\documentstyle[12pt]{article}

\def\sdspace{\baselineskip = .25in}

\def\VEV#1{\left\langle #1\right\rangle}

\begin{document}
\title{Family Symmetry, Gravity, and the\\ Strong CP Problem\thanks{Supported
in part by Department of Energy Grant
\#DE-FG02-91ER406267}}

\author{{\bf K.S. Babu} and {\bf S.M. Barr}\\ Bartol Research Institute\\
University of Delaware\\ Newark, DE 19716}

\date {BA-92-79}
\maketitle
\begin{abstract}
We show how in a class of models
Peccei--Quinn symmetry can be realized as an automatic
consequence of a gauged $U(1)$ family symmetry.  These models
provide a solution to the strong CP problem either via a massless
$u$--quark or via the DFSZ invisible axion.  The local family symmetry
protects against potentially large corrections to $\overline{\theta}
$ induced by quantum gravitational effects.  In a supersymmetric
extension, the `$\mu$--problem'
is shown to have a natural solution in the
context of gravitationally induced operators.  We also present a
plausible mechanism which can explain
the inter--generational mass hierarchy in such a context.
\end{abstract}
\newpage
\sdspace

{\bf I.}  Global symmetries have lately come under suspicion.
There are arguments that quantum gravitational effects violate
global symmetries and will induce in the effective low--energy theory
all possible operators that respect the local symmetries of the theory.$
^1$
The magnitudes of the coefficients of these operators are expected to be
set by the appropriate powers of the Planck scale, but beyond that no
quantitative statements are possible at present.  In superstring theory
there are firmer arguments against the possibility of exact global
continuous symmetries.$^2$

This suspicion has led various authors to reevaluate theoretical ideas
that involve global symmetries, such as Peccei--Quinn symmetries,$^3$ CP,
$^4$
baryon$^5$ and lepton numbers$^6$ and cosmological texture.$^7$
The idea has been to see if such global
symmetries can arise as an automatic consequence of local symmetries.

In this paper we will reexamine Peccei--Quinn symmetry.$^8$  Peccei--Quinn
symmetry can be realized in two ways: either in the Wigner--Weyl way
which leads to a massless quark (or quarks) usually taken to be the
$u$--quark; or in the Nambu--Goldstone way giving rise to an axion.$^9$  We
will first show how local family symmetries can lead to a $u$--quark
light enough to solve the strong CP problem.  Then we will show how, in
a closely related fashion, very simple DFSZ axion models$^{10}$ can arise from
local family symmetry realizing an idea of Wilczek.$^{11}$  Generalization to
grand unification and supersymmetry will be presented.  A natural
solution to the `$\mu$--problem' in SUSY models is found in the
context of quantum gravity induced operators.  We also construct a
scheme in this context which can explain the
inter--generational mass--hierarchy.

We adopt the philosophy in this paper that the operators presumably
induced by quantum gravitational effects are not further highly
suppressed (beyond the powers of $M_{\rm Pl}$ expected on dimensional
grounds).  There are no obvious small parameters which would lead us to
expect such a suppression.  We therefore take the dimensionless
coefficients in front of these operators to be of order one.

{\bf II.} {\it Massless $u$--quark:}  Let us assume that the low
energy theory looks like the standard model, but that there is secretly
a $U(1)$ local symmetry broken at some high scale.  Call this
$U(1)^{\prime}$.  Let us assign the $U(1)^{\prime}$ charges of the
standard model fields as shown in Table I.  The $U(1)^{\prime}$ charges
of the right--handed quarks and leptons ($D_L^{c(i)}, U_L^{c(2,3)},
l_L^{c(i)})$ are determined by the usual standard model Yukawa couplings
in terms of the charges of the other fields ($i=$ family index).  Note
that an exception is the right--handed $u$--quark since we do not want
its usual Yukawa coupling to be present.  (Choosing the
$U(1)^{\prime}$ charge of the left-handed quark doublet $Q_L^{(1)}$
could
also lead to a massless $u$--quark, but in this case there will be no
Cabibbo mixing involving the first family.)  Its $U(1)^{\prime}$ charge
we require to differ from those of the right--handed $c$ and $t$ by an
amount $\Delta$.  Now, if we suppose that the unknown and presumably
heavy fermions that are required to exist for anomaly freedom are all
color singlets, we find from the $SU(3)_C^2 \times U(1)^{\prime}$
anomaly condition
$$0=6a+3(-a+q)+2(-a-q)+(-a-q+\Delta)~,$$
or $\Delta=0$.  In other words, the $U(1)^{\prime}$ fails to distinguish
the $u$--quark and protect it from having a tree--level mass.  There are two
simple ways out.

(1) One can allow exotic colored fermions to contribute to the
anomaly.  The simplest model of this type is the following:  Have the
$U(1)^{\prime}$ charge, $Q^{\prime}$, vanish for all standard model
fermions except $u_L^c$ which has $Q^{\prime}=1$, and add a pair of
quarks with $SU(3)_C \times SU(2)_L \times U(1)_Y \times U(1)^{\prime}$
quantum numbers $(3,1,-2/3,0)_L+(\overline{3},1,+2/3,-1)_L$.  Note that
these have exotic electric charges--the quark has electric charge
$minus$ $2/3$.  All anomalies cancel.  A Higgs field,
$S(1,1,0,1)$ can give Dirac mass to the exotic pair of quarks.  A
dimension--4 mass term for $u$ is forbidden; however, a term
$${f \over {M_{\rm Pl}}}\left(Q_Lu_L^c\varphi S^*\right)$$
would be expected to be induced by quantum gravity.  In order to solve
the strong CP problem $m_u$ should be less than about $10^{-9}$ of its
usually assumed value of $5$ to $10$ MeV, so that $f\VEV{S}/M_{\rm Pl}
\stackrel{_<}{_\sim} 10^{-14}$.  One therefore expects the exotic quarks
to have mass of order $\VEV{S} \sim (100~TeV)/f$.  Unfortunately such
exotically charged quarks (unless they could be inflated away) would run
afoul of terrestrial searches for anomalously charged matter.

(2) To us a more interesting possibility is that we modify the
standard model by having two Higgs doublets.  Suppose $\varphi$ couples
as usual to the $c$ and $t$ quarks (but not $u$), and $\varphi^{\prime}$
couples as usual to charge $-1/3$ quarks and leptons.  (In our notation
$\varphi$ and $\varphi^{\prime}$ both have hypercharge = $+1/2$.)
The $U(1)^{\prime}$ quantum
numbers of the various fields are now given as in Table II.  The same
exercise as before shows that the $SU(3)_C^2 \times U(1)^{\prime}$
anomaly cancels if
\begin{equation}
\Delta = 3(q-q^{\prime})~~.
\end{equation}
So the $u$ quark mass is forbidden by $U(1)^{\prime}$ only if $\varphi$
and $\varphi^{\prime}$ have different $U(1)^{\prime}$ charges.  This
means that the $\varphi^{\dagger}\varphi^{\prime}$  and the
$(\varphi^{\dagger}\varphi^{\prime})^2$ terms are also forbidden.  There
is then the danger of a weak scale axion.  If that happens, there would
be overkill: two accidental PQ symmetries would result, the usual
Weinberg--Wilczek one realized in the Nambu--Goldstone way, and another
PQ symmetry that takes only $u^c \rightarrow e^{i\alpha}u^c$ realized in
the Wigner--Weyl way.  However, the axion is easily avoided if the
singlet field $S$ needed to break $U(1)^{\prime}$ above the weak scale
has a coupling
\begin{equation}
\varphi^{\dagger}\varphi^{\prime}S^n/M_{\rm Pl}^{n-2}~~.
\end{equation}
Here $n$ is some appropriate integer determined by the
$U(1)^{\prime}$ charges.  If $n=1$ or $2$, eq. (2)
is a part of the renormalizable Lagrangian, for $n > 2$, eq. (2) will
only be induced by quantum gravitational effects.  From eq. (1) and (2),
one sees that there will also be a possible term induced by gravity
\begin{equation}
Q_Lu^c\varphi(S^*)^{3n}/M_{\rm Pl}^{3n}~~.
\end{equation}
We require $\VEV{S}^n/M_{\rm Pl}^{n-2} \sim M_W^2$ to avoid either an
axion or the disruption of the gauge hierarchy.  Then
$$m_u \sim M_W\left(M_W^2/M_{\rm Pl}^2\right)^3$$
which is certainly small enough to solve the strong CP problem.  We
should note also that the usual color--instanton contribution to $m_u$
will still arise as in the standard model$^{12}$ as these arise from operators
like $Q_Lu^c\varphi^{\prime}\left(\varphi^{\dagger}\varphi^{\prime}
\right)^2$ which are allowed by $U(1)^{\prime}$, not surprisingly since
$U(1)^{\prime}$ is constructed to have no color anomaly.

It is worth showing that a set of extra fermions that is not too wild
can be found that ensures cancellation of all gauge anomalies (see Table
III).  These extra fermions are singlets under color and $SU(2)_L$.
The $SU(3)^2 \times U(1)^{\prime}$ and $SU(2)^2 \times U(1)^{\prime}$
and gravity $\times U(1)^{\prime}$ anomaly cancellation imply
$\Delta = 3(q-q^{\prime}),~b=-3a,$ and $q^{\prime}=-b$, respectively (see
Tables II and III).  The $U(1)_Y^2 \times U(1)^{\prime}$ anomaly then
automatically cancels, and the $U(1)_Y \times U(1)^{\prime^2}$ and
$U(1)^{\prime^3}$ anomaly conditions are respectively
\begin{eqnarray}
2yz(x_1-x_2-z) &=& 12(q-q^{\prime})^2 \nonumber \\
3(x_1+x_2)z(x_1-x_2-z) &=& 18(q-q^{\prime})^2(5q^{\prime}-q)~~. \nonumber
\end{eqnarray}
These give $x_1+x_2=y(5q^{\prime}-q)$ and
\begin{equation}
z^2+(x_2-x_1)z+{6 \over y}(q-q^{\prime})^2 = 0~~.
\end{equation}
The heavy fermions $\psi_1$ and $\psi_2$ have electric charge $y$ which
has to be different from zero.  The simplest choice is $y=1$, so that
they are heavy leptons of the ordinary type.  $\psi_1$ and $\psi_2$ can
both get mass from a scalar $S$ of charge $Q^{\prime}=z/m$ if the mass
terms are of the form $\psi_1^c \psi_1(S^*)^m+\psi_2^c\psi_2S^m$.  Note
that for $m=1$, these mass terms are part of the renormalizable
Lagrangian.  It would be nice if the same scalar, $S$, coupled to
$\varphi^{\dagger}\varphi^{\prime}$ as in eq. (2), and prevented an
axion.  The scalar $S$ from eq. (2) we see has $U(1)^{\prime}$ charge
$(q-q^{\prime})/n$.  So if $z/m=(q-q^{\prime})/n$, eq. (4) gives
$(x_2-x_1)=(q^{\prime}-q)\left({6 \over y}{n \over m}+ {m \over n}
\right)$, which together with $(x_2+x_1)=y(5q^{\prime}-q)$ gives nice rational
values for the $U(1)^{\prime}$ charges of the extra fermions.  Note that
the masses of these extra fermions are expected to be of order
$M_{\rm Pl}(M_W/M_{\rm Pl})^{2m/n}$.  The special choice $m=1,~n=2$,
which guarantees that the axion mass as well as $\psi_{1,2}$ masses
arise at tree level would imply that the masses of $\psi_{1,2}$ are of
order $M_W$.

It should be pointed out that ``discrete gauge symmetries''$^{13}$ that survive
at low energy are useless in making $m_u=0$ as the same would also lead to
weak axions. If a $\varphi^{\dagger}\varphi^{\prime}$ term is present, it means
any residual $Z_N$ must be such that $q-q^{\prime}=0$ mod $N$.  But then
$\Delta=3(q-q^{\prime}) = 0$ mod $N$ and the $Z_N$ fails to distinguish
the $u$ from the $c$ and $t$.

{\bf III.} {\it Family symmetry and automatic DFSZ axions:}  From the
foregoing we see that there is a quite natural link between gauged family
symmetries and axions.  We tried to impose a $U(1)^{\prime}$ symmetry
that distinguished $u$ from $c$ and $t$ (i.e., a simple family
symmetry) and found that we were led to consider two--Higgs--doublet
models where the same family symmetry distinguished the two Higgs
doublets and prevented a $\varphi^{\dagger}\varphi^{\prime}$ term.  We will now
show how to exploit this to construct very simple DFSZ axion models.

Consider the very model we discussed in the last section where the
fermion content is displayed in Tables II and III.  Let there be $two$
singlet scalars $S$ and $T$.  $T$ we give $Q^{\prime}=\Delta=3(q-
q^{\prime})$ (the latter equality following, again, from the $SU(3)^2
\times U(1)^{\prime}$ anomaly freedom).  $S$ we give $Q^{\prime} =
{1 \over n}(q-q^{\prime})$.  Then there are the terms
\begin{eqnarray}
{}~~~~~~~~~~~~~~~~~~~~~~~~~~~~~~~~~
\varphi^{\dagger}\varphi^{\prime}\left(S^n/M_{\rm Pl}^{n-2}\right)
{}~~~~~~~~~~~~~~~~~~~~~~~~~~~~~~~~~~~~~~(5a)\nonumber \\
{}~~~~~~~~~~~~~~~~~~~~~~~~~~~~~~~~~
Q_L u^c\varphi \left(T^*/M_{\rm Pl}\right)
{}~~~~~~~~~~~~~~~~~~~~~~~~~~~~~~~~~~~~~~~(5b)
\nonumber \\
{}~~~~~~~~~~~~~~~~~~~~~~~~~~~~~~~~~
\left(T^*S^{3n}/M_{\rm Pl}^{3n-3}\right)
{}~~~~~~~~~~~~~~~~~~~~~~~~~~~~~~~~~~~~~~~(5c)
\nonumber
\end{eqnarray}
(There is also the term $Q_L u^c \varphi^{\prime}(\varphi^{\dagger}
\varphi^{\prime})^2$ discussed earlier, but this is negligible.)
Let us ignore the term (5c) for the moment since for the cases of
interest it will be of high order in $(1/M_{\rm Pl})$.  We are
interested, now, in solving the strong CP problem via an axion rather
than a massless or very light $u$ quark.  So we choose $\VEV{T}$ to be
large enough so that $m_u$ arising from (5b) is about $5$ to $10$ MeV.
Thus $\VEV{T} \stackrel{_>}{_\sim}10^{15}~GeV$.  Now there are (if we
neglect (5c)) two $U(1)$ symmetries (besides $U(1)_Y$) to consider:
the local symmetry $U(1)^{\prime}$ and the anomalous (and accidental)
global $U(1)$ symmetry that takes $T \rightarrow e^{i \alpha}T$ and
$u^c \rightarrow e^{i \alpha}u^c$.  We will call the latter symmetry
$U(1)_T$.  These are both broken by $\VEV{S}$ and $\VEV{T}$.  So the
$U(1)^{\prime}$ gauge boson will become massive, and there will also be
an axion.  What is $f_a?$  Assume $\VEV{T} \gg \VEV{S}$.  Then
$\VEV{T}$ will break $U(1)^{\prime} \times U(1)_T$ down to a global
$U(1)$ and the $U(1)^{\prime}$ gauge boson will eat the phase of $T$.
(This is the so--called 'tHooft mechanism.)  The residual global $U(1)$,
which we will call $U(1)_{PQ}$, will be that linear combination of
$U(1)^{\prime}$ and $U(1)_T$ under which $T$ is neutral.  The
$U(1)_{PQ}$ charges of all the fermions will be the same as their
$U(1)^{\prime}$ charges, $except$ for $u^c$.  From (5b) and the fact
that $Q_{PQ}(T)=0$ we see that $u^c$ has the same PQ charge as $c^c$ and
$t^c$.  But this PQ symmetry is then just the familiar DFSZ kind of
$U(1)$.  It gets broken by $\VEV{S}$, so that for cosmological and
astrophysical reasons $10^{10}~GeV \stackrel{_<}{_\sim} \VEV{S}
\stackrel{_<}{_\sim}10^{12}~GeV$.

The coefficient of $\varphi^{\dagger}\varphi^{\prime}$ will be of
order $M_W^2$ if $n$ is chosen to be $4$ (see eq. (5a)).  Recall that in
the usual DFSZ models $n=2$ (or $1$), which would require
fine--tuning the coefficient of $\varphi^{\dagger}\varphi^{\prime}$ to
be of order $M_W^2$.

Up to this point we
have neglected the term (5c).  This term explicitly violates the
Peccei--Quinn symmetry and therefore, as emphasized in ref. (3),
contributes to $\overline{\theta}$.  One expects that this contribution
will be
$$ \overline{\theta} \sim (m_\pi^2f_\pi^2)^{-1}{{\VEV{T^*}\VEV{S}^{3n}}
\over {M_{\rm Pl}^{3n-3}}}~.$$  Using $(S/M_{\rm Pl})^n \sim
(M_W/M_{\rm Pl})^2$ (from (5a)), $\VEV{T}/M_{\rm Pl} \sim m_u/M_W$ (from
(5b)), one finds $$\overline{\theta} \sim 10^{-26}~~.$$  This relation
depends on the power $3n$ that appears in (5c).  The  $3$ comes from the
$SU(3)^2 \times U(1)^{\prime}$ anomaly condition $\Delta =
3(q-q^{\prime})$, and depends on the charge assignment of the quarks
under the $U(1)^{\prime}$ family group.

{\bf IV.} {\it GUT embedding:}  The family group we have used is
somewhat peculiar: it distinguishes the $u^c$ quark only.  One might ask
whether more general family groups are possible, including ones that
would commute with grand unified gauge groups.  To see that this is
indeed the case we will describe a simple $SU(5) \times U(1)^{\prime}$
example.  Consider three families of quarks and leptons, each contained
in a  $10 + \overline{5}$ of $SU(5)$.  Let $i=1,2,3$ be the
family index.  Assign to the fermion representations $10_L^{(i)}$
$U(1)^{\prime}$ charges $a+p_i\Delta$, to the $\overline{5}_L^{(i)}$
$U(1)^{\prime}$ charges $b+q_i\Delta$, and to the Higgs representations
$5_H$ and $\overline{5}_H$ $U(1)^{\prime}$ charges $q=-2a$ and
$-q^{\prime}=-(a+b)$ respectively.  If the integers $p_i,~q_i$ all
vanished, then the usual Yukawa couplings $10^{(i)}\overline{5}^{(j)}
\overline{5}_H$ and $10^{(i)}10^{(j)}5_H$ would be allowed, and freedom
from the $SU(5)^2 \times U(1)^{\prime}$ anomaly would imply $9a+3b=0$.
One recognizes the resulting $U(1)^{\prime}$ as just being contained in
$SO(10)$:  $SU(5) \times U(1)^{\prime} \subset SO(10)$.  (Of course,
other fermions need to be added, here $\nu_R$'s would do, to cancel the
$U(1)^{\prime^3}$ anomaly.)  But if the $p_i,~q_i$ are not all zero,
$U(1)^{\prime}$ is a family symmetry and forbids certain $d=4$ Yukawa
terms.  New $SU(5)$--singlet fermions are in some cases required for
$U(1)^{\prime^3}$ anomaly cancellation, but their presence will not
affect our results.
As before, we introduce $SU(5)$ singlet fields $S$ and $T$ with
$U(1)^{\prime}$ charges $(q^{\prime}-q)/n=(3a+b)/n$ and $\Delta$
respectively.  Then the couplings
\begin{eqnarray}
{}~~~~~~~~~~~~~~~~~~~~~~~~~~~
(5_H \overline{5}_H)S^n/M_{\rm Pl}^{n-2} ~~~~~~~~~~~~~~~~~~~~~~~~~~~~~~~~(6a)
\nonumber \\
{}~~~~~~~~~~~~~~~~~~~~~~~~~~~
(10^{(i)}\overline{5}^{(j)}\overline{5}_H)(T^*/M_{\rm Pl})^{p_i+q_j}~~~~~~
{}~~~~~~~~~~~~~~~~~~~~~~~~~~(6b) \nonumber \\
{}~~~~~~~~~~~~~~~~~~~~~~~~~~~(10^{(i)}10^{(j)}5_H)(T^*/M_{\rm Pl})^{p_i+p_j}
{}~~~~~~~~~~~~~~~~~~~~~~~~~~~~~~~~(6c)\nonumber
\end{eqnarray}
should be induced by gravity.  That is, the missing $d=4$ Yukawa
interactions appear in the effective low energy theory suppressed by
appropriate power of $\VEV{T}/M_{\rm Pl}$.

Now, the $SU(5)^2 \times U(1)^{\prime}$ anomaly tells us that
$$\sum_{i} 3(a+p_i\Delta)+\sum_{i}(b+q_i\Delta) = 0~,$$ or
$$\Delta = -(3a+b) {3 \over {[3\Sigma p_i+\Sigma q_i]}}$$ where
$(3a+b)=(q^{\prime}-q)$.  Therefore one has a term allowed by
$U(1)^{\prime}$ and, hence presumably induced by quantum gravity
\begin{eqnarray}
{}~~~~~~~~~~~~~~~~~~~~~~~~~T^{[3\Sigma p_i+\Sigma q_i]}S^{3n}/M_{\rm Pl}^{3n-
[3\Sigma p_i+
\Sigma q_i]-4}~.~~~~~~~~~~~~~~~~~~~~~~~~~~~(7)~ \nonumber
\end{eqnarray}
As in the previous example, this generates a $\overline{\theta}$ which
is very small since it is proportional to $(\VEV{S}/M_{\rm Pl})^{3n}
\sim (M_W^2/M_{\rm Pl}^2)^3$.  However, it should be noted that a problem
would arise if the integer $[3\Sigma p_i+\Sigma q_i]$ were a multiple
of $3$.  Then an operator that is the $3$rd root of eq. (7) would be
induced by gravity that would in general give too large a $\overline{
\theta}$.  Moreover, in that case an operator $5_H \overline{5}_H
T^{[3\Sigma p_i+ \Sigma q_i]/3}$ would endanger the gauge hierarchy.
The same danger exists if $[3\Sigma p_i+\Sigma q_i]$ and $3n$ have any
common divisor.  Note the significant fact that if all the $p_i$ are equal
and all $q_i$ are equal, then $[3\Sigma p_i+\Sigma q_i]$ is a multiple
of $3$, so that to avoid the aforementioned problems $U(1)^{\prime}$
must truly be a family symmetry, i.e., it should distinguish among
families.

{\bf V.} {\it Supersymmetric extension:}  All of the above
considerations apply to supersymmetric models as well, with certain
significant changes.  Consider, for example, the supersymmetric version
of the model we just discussed.  (6a)-(6c) are then to be interpreted as
terms in the superpotential (with one more power of $M_{\rm Pl}$ in the
denominator of (6a) to make the dimensions come out right).  Note that
$R$--parity violating couplings such as $10^{(i)}\overline{5}^{(j)}
\overline{5}^{(k)}$ and $10^{(i)}\overline{5}_H\overline{5}_H$ are forbidden
by the $U(1)$ family symmetry.  Eq. (6a) implies
that the $\mu$ parameter is
$$ \mu \sim \VEV{S}^n/M_{\rm Pl}^{n-1}$$ or $$(\VEV{S}/M_{\rm Pl})^n
\sim M_W/M_{\rm Pl}$$ (instead of $M_W^2/M_{\rm Pl}^2$ as in the
non--supersymmetric case).  So for $10^{10}~GeV \stackrel{_<}{_\sim}
\VEV{S} \stackrel{_<}{_\sim} 10^{12}~GeV$, choosing
$n=2$ would `explain' why the $\mu$ parameter is of the weak scale (and
not the Planck scale).  The anomaly
condition, including now the effect of $\overline{5}_H,~ 5_H$ is
$$\Delta = -(3a+b) {2 \over {[3\Sigma p_i + \Sigma q_i]}}$$
and the gravity induced term contributing to $m_a$ is
$$T^{[3 \Sigma p_i+ \Sigma q_i]} S^{2n}/M_{\rm Pl}^{2n+[3 \Sigma p_i
+ \Sigma q_i]-3}~.$$
We see that to make $\overline{\theta} \stackrel{_<}{_\sim} 10^{-9}$
requires $(\VEV{T}/M_{\rm Pl})^{[3 \Sigma p_i+\Sigma q_i]}
\stackrel{_<}{_\sim} 10^{-55}$, if $[3\Sigma p_i+\Sigma q_i]$ and
$2n=4$ have no common divisor.  This is obviously a rather stringent
condition, but in other models the condition would be different.

A further remark is in order as regards the `doublet--triplet
mass--splitting' in SUSY $SU(5)$.  Since the Higgs doublet has a mass of
order $M_W$ in our scheme, question may be raised as to the origin of
the superheavy mass of its color--triplet partner.  Other known
mechanisms, such as the `missing partner mechanism' are compatible with
our scheme and could give superlarge mass to the
color triplets.

{\bf VI.} {\it Fermion mass hierarchy:}  It is possible in the present
context to have a natural explanation of the inter--generational mass
hierarchy.  Take for example the SUSY model of the previous section.
The quark and lepton masses arise from eq. (6b) and (6c).  Suppose we
choose the $U(1)^{\prime}$ charges such that $p_i=(2,1,0)$ and
$q_i=(1,1,1)$ in the notation introduced earlier.  Then if
$x \equiv \VEV{T}/M_{\rm Pl}
\sim \sqrt{m_c/m_t} \simeq 1/15$, and tan$\beta$ (the ratio of the
Higgs vacuum expectation values) $\simeq 3$, the fermion mass matrices
will have the form
\begin{eqnarray}
M_{u} \sim v{\rm sin}\beta \left(\matrix{x^4 & x^3 & x^2 \cr x^3 & x^2 & x \cr
x^2 & x & 1}\right)~~;~~ M_{d,l} \sim v {\rm cos}\beta \left(\matrix{
x^3 & x^3 & x^3 \cr x^2 & x^2 & x^2 \cr x & x & x}\right) ~,\nonumber
\end{eqnarray}
where $v \simeq 175~GeV$.  (Numbers of order one multiplying various
entries in the matrices have been dropped.)  Such
matrices give a nice hierarchy of masses as well as mixing angles.  Note
that the mass ratios in the up--sector are smaller by a factor $x$
relative to those in
the down sector, in agreement with observations.
We have not attempted to reconcile the strong CP problem simultaneously
with the mass hierarchy, but models which accomplish both are not
inconceivable.

{\bf VII.} {\it Conclusion:}  We found that a local family symmetry can
make $m_u$ light enough to solve the strong CP problem.  However, if
this symmetry has a residue at low energy that is a ``discrete gauge
symmetry'' there is the tendency to get a weak axion as well.
We also found that a DFSZ kind of Peccei--Quinn symmetry can arise very
naturally as a consequence of local $U(1)$ family symmetries.  This
approach has several appealing features: (i) there is a direct
connection between the scale of Peccei--Quinn breaking and the value of
the $\mu$ parameter; (ii) the $\mu$ parameter arises as a result of
gravitationally induced terms and its smallness is in some sense
explained; (iii) the choice of family group, the value of $f_a$, the
value of $\overline{\theta}$ and the size of certain light quark and
lepton masses are linked together.
Of course, these models suffer the great defect of all the DFSZ models
that they are hard (impossible?) to test.  But perhaps the ideas
suggested here will allow further progress on the idea of family
symmetry.
\newpage
\section*{References}
\begin{enumerate}
\item For reviews, see for example,
T. Banks, Physicalia {\bf 12}, 19 (1990); J. Preskill, Caltech
Preprint CALT-68-1819 (1992).
\item T. Banks, L. Dixon, D. Friedan and E. Martinec, Nucl. Phys. {\bf
B299}, 613 (1988).
\item M. Kamionkowski and J. March-Russel, Phys. Lett. {\bf B282}, 137 (1992);
R. Holman et. al., Phys. Lett. {\bf B282}, 132 (1992);
S.M. Barr and D. Seckel, Phys. Rev. {\bf D 46}, 539 (1992).
\item M. Dine, R. Leigh and D. MacIntire, Phys. Rev. Lett. {\bf 69}, 2030
(1992); K. Choi, D. Kaplan and A. Nelson, San Diego Preprint (1992);
J.E. Kim, Seoul National University Preprint SNUTP 92-52 (1992).
\item G. Gelmini and R. Holman, Santa Barbara Preprint NSF-ITP-92-101
(1992).
\item E.Kh. Akhmedov, Z. Berezhiani and G. Senjanovic, Phys. Rev. Lett.
{\bf 69}, 3013 (1992);
E.Kh. Akhmedov, Z. Berezhiani, R.N. Mohapatra and G. Senjanovic,
Maryland Preprint UMD-HEP-93-020 (1992);
I. Rothstein, K.S. Babu and D. Seckel, Bartol Preprint BA-92-77
(1992).
\item R. Holman et. al., Phys. Rev. Lett. {\bf 69}, 1489 (1992);
M. Kamionkowski and J. March-Russel, $ibid.,$ 1485.
\item R.D. Peccei and H. Quinn, Phys. Rev. Lett. {\bf 38}, 1440 (1977);
Phys. Rev. {\bf D16}, 1791 (1977).
\item S. Weinberg, Phys. Rev. Lett. {\bf 40}, 223 (1978); F. Wilczek, $
ibid$., 279.
\item M. Dine, W. Fischler and M. Srednicki, Phys. Lett. {\bf 104B}, 199
(1981); A.P. Zhitnitskii, Sov. J. Nucl. Phys. {\bf 31}, 260 (1980).
\item F. Wilczek, Phys. Rev. Lett. {\bf 49}, 1549 (1982).
\item D. Kaplan and A. Manohar, Phys. Rev. Lett. {\bf 56}, 2004 (1986);
K. Choi, C.W. Kim and W.K. Sze, $ibid$., {\bf 61}, 794 (1988).
\item L. Krauss and F. Wilczek, Phys. Rev. Lett. {\bf 62}, 1221 (1989);
J. Preskill and L. Krauss, Nucl. Phys. {\bf B341}, 50 (1990); L. Ibanez
and G. Ross, Phys. Lett. {\bf 260B}, 219 (1991).
\end{enumerate}

\newpage

{\bf Table I.}  $U(1)^{\prime}$ charges of the standard model fermions and
the Higgs doublet.
\vskip .5in
\begin{center}
\begin{tabular}{|c|c|c|c|c|c|c|}\hline
$Q_L^{(i)}$ & $D_L^{c(i)}$ & $U_L^{c(2,3)}$ & $u_L^c$ & $L^{(i)}$ &
$l^{c(i)}$ & $\varphi$ \\
\hline
$a$ & $(-a+q)$ & $(-a-q)$ & $(-a-q+\Delta)$ & $b$ & $(-b+q)$ & $q$ \\
\hline
\end{tabular}
\end{center}
\vskip 1in
{\bf Table II.}  $U(1)^{\prime}$ charge assignment in the two Higgs doublet
model.
\vskip .5in
\begin{center}
\begin{tabular}{|c|c|c|c|c|c|c|c|}\hline
$Q_L^{(i)}$ & $D_L^{c(i)}$ & $U_L^{c(2,3)}$ & $u_L^c$ & $L^{(i)}$ &
$l^{c(i)}$ & $\varphi$ & $\varphi^{\prime}$ \\
\hline
$a$ & $(-a+q^{\prime})$ & $(-a-q)$ & $(-a-q+\Delta)$ & $b$ & $(-b+
q^{\prime})$ & $q$ & $q^{\prime}$ \\
\hline
\end{tabular}
\end{center}
\vskip 1in
{\bf Table III.}  Hypercharge and $U(1)^{\prime}$ quantum numbers of heavy
leptons needed for anomaly cancellation.
\vskip .5in
\begin{center}
\begin{tabular}{|c|c|c|c|c|}\hline
{}~ & $\psi_{1L}$ & $\psi_{1L}^c$ & $\psi_{2L}$ & $\psi_{2L}^c$ \\
\hline
$Y$ & $y$ & $-y$ & $y$ & $-y$ \\
$Q^{\prime}$ & $x_1$ & $-x_1+z$ & $x_2$ & $-x_2-z$ \\
\hline
\end{tabular}
\end{center}

\end{document}